\documentclass[journal,comsoc]{IEEEtran}
\usepackage[T1]{fontenc}
\usepackage{easyReview}
\usepackage{multirow}
\usepackage{comment}
\usepackage{hyperref}
\usepackage{cite}
\usepackage{dblfloatfix}
\usepackage{graphicx}
\usepackage{amsmath}
\usepackage{cleveref}
\usepackage{caption}
\usepackage[caption=false,font=footnotesize]{subfig} 
\usepackage{lipsum}

\crefname{figure}{fig.}{figs.}           

\begin{document}

\title{Dynamic DV-QKD Networking in Trusted-node-free Software-Defined Optical Networks}
%
%
%

\author{{Obada~Alia, Rodrigo~Stange~Tessinari, Emilio~Hugues-Salas, George~T~Kanellos, Reza~Nejabati, Dimitra~Simeonidou}

\thanks{O.~Alia G.~Kanellos, R.~Nejabati and D.~Simeonidou are with the High Performance Network Group School of Computer Science, Electrical \& Electronic Engineering and EngineeringMaths (SCEEM), University of Bristol BS8 1TH, Bristol, UK, e-mail: (obada.alia@bristol.ac.uk).}
\thanks{R.~S.~Tessinari is with Toshiba Europe Limited, Cambridge Research Laboratory, 208 Science Park, Milton Road, Cambridge CB4 0GZ, UK. Previously with the High Performance Network Group at the University of Bristol}
\thanks{E.~Hugues-Salas is with BT Labs, Adastral Park, Martlesham Heath, UK. Previously with the High Performance Network Group at the University of Bristol}
}

\markboth{Journal of Lightwave Technology, VOL. XX, No. X, 2022.}
{}

\maketitle

\begin{abstract}

We demonstrate for the first time a four-node trusted-node-free metro network configuration with dynamic discrete-variable quantum key distribution DV-QKD networking capabilities across four optical network nodes. The network allows the dynamic deployment of any QKD link between two nodes of the network, while a QKD-aware centralised software-defined networking (SDN) controller is utilised to provide dynamicity in switching and rerouting. The feasibility of coexisting a quantum channel with carrier-grade classical optical channels where both the quantum and classical channels are in the C-band over field-deployed metropolitan networks and laboratory-based fibres (<10km) is experimentally explored in terms of achievable quantum bit error rate, secret key rate as well as classical signal bit error rate. Moreover, coexistence analysis over multi-hops configuration using different switching scenarios is also presented. The secret key rate dropped 43\% when coexisting one classical channel with 150~GHz spacing from the quantum channel for multiple links. This is due to the noise leakage from the Raman scattering into the 100~GHz bandwidth of the internal filter of the Bob DV-QKD unit. When coexisting four classical channels with 150~GHz spacing between quantum and the nearest classical channel, the quantum channel deteriorates faster due to the combination of Raman noise, other nonlinearities and high aggregated launch power causing the QBER value to exceed the threshold of 6\% leading the SKR to reach a value of zero bps at a launch power of 7~dB per channel. Furthermore, the coexistence of a quantum channel and six classical channels through a field-deployed fibre test network is examined.

\end{abstract}


%
\IEEEpeerreviewmaketitle

\section{Introduction}

\IEEEPARstart{Q}{uantum} communication networks are evolving from relayed point-to-point quantum communication links to multi-user quantum networks with partial or full connectivity \cite{mehic2020quantum, {pirandola2020advances}}. Although the current quantum networks are mainly implementing Quantum Key Distribution (QKD) protocols which is the most mature quantum communication application \cite{gisin2002quantum}, the future of quantum communication networks is to provide a worldwide connectivity via complex networks with a seamless interconnection between multiple nodes to enable applications beyond QKD such as blind and distributed quantum computing \cite{van2016path}. For such networks to reach their maximum capabilities, fully dynamic quantum network solutions \cite{tessinari2019field} need to be deployed to eventually overcome the established practice of relayed point-to-point quantum links and allow efficient resource sharing across the networks.

Since Dynamic QKD would rely on the need for deployment of low-loss switching and routing elements in the QKD channel, which may compromise its performance. The concept leads to a trade-off between QKD performance in terms of secret key rate (SKR) and increased network functionality with improved capabilities to maximise resource usage optimisation.  However, this may be particularly appealing for optical networks in dense metropolitan regions of big cities, with the presence of large number of nodes likes with short fibre length (<10km) \cite{MAJUMDAR20195}. In addition, allowing the coexistence of WDM classical and quantum channels over the same fibre in metropolitan optical networks with the ability to switch between different nodes providing multi-hops connectivity \cite{MAJUMDAR20195} could be further beneficial for small scale quantum networks alleviating the need for additional dark fibres deployment. However the concept only applies for short fibre lengths due to the limited power budget of the QKD systems \cite{eraerds2010quantum, patel2012coexistence} and the coexistence in such fibres is extremely challenging due to the higher impact of the nonlinear effects on the quantum channel and therefore for longer distance fibres, quantum repeaters should be employed.

QKD utilises quantum mechanics to generate information-theoretic keys that can be used to encrypt and decrypt key information using classical symmetric-key algorithms such as One Time pad or Advanced Encryption Standard (AES).  However, for QKD networks to become functional for the future Internet use cases such as 5G, the integration of QKD with classical optical networking infrastructure and adaptation the dynamic environment to the metro and edge part of optical networks are crucial; hence dynamic QKD could be considered as a potential security candidate for such networks \cite{zavitsanos2020qkd}. 

\begin{table*}[!t]
\centering
\caption{QKD networks properties}
\label{tab:QKD_networks}
\begin{tabular}{@{}cccccccc@{}}
\hline
\hline
\multicolumn{1}{c}{QKD Network}  & \multicolumn{1}{c}{Number of Nodes}   & \multicolumn{1}{c}{Trusted nodes} & \multicolumn{1}{c}{Coexistence}  & \multicolumn{1}{c}{SDN} & \multicolumn{1}{c}{Topology} & \multicolumn{1}{c}{Key references}                                                     
\\ \hline
\multicolumn{1}{c}{DARPA QKD Network} & 10 & Yes & No & No & Mesh &\cite{elliott2005current} \\
\multicolumn{1}{c}{SECOQC QKD Network} & 6 & Yes & No & No & Mesh & \cite{peev2009secoqc,poppe2008outline} \\
\multicolumn{1}{c}{Tokyo QKD Network}  & 6 & Yes & No & No & Mesh & \cite{sasaki2011field} \\
\multicolumn{1}{c}{China QKD Network} & 109 & Yes & No & No & Mesh &\cite{chenintegrated}\\ 
\multicolumn{1}{c}{Bristol Entanglement-based QKD Network } & 8 & No & No & No & Full Mesh &\cite{joshi2020trusted} \\
\multicolumn{1}{c}{Cambridge-Ipswich QKD Network} & 5 & Yes & Yes & No & Bus &\cite{wonfor2019field} \\
\multicolumn{1}{c}{Cambridge QKD Network} & 3 & Yes & Yes & No & Ring &\cite{dynes2019cambridge} \\
\multicolumn{1}{c}{Madrid QKD Network} & 3 & No & Yes & Yes & Ring &\cite{martin2019madrid, aguado2019engineering}\\
\multicolumn{1}{c}{Bristol Dynamic QKD Network [This Work]} & 4 & No & Yes & Yes & Mesh &\cite{tessinari2019field}\\ 
\hline
\end{tabular}
\end{table*}

Table \ref{tab:QKD_networks} summarises the main features of the major QKD networks. Although the first field-trial of a QKD network was the DARPA QKD demonstrated in 2004 \cite{elliott2005current}, it did not provide any practical implementation of the QKD system. The SECOQC was a subsequent demonstration of a QKD network aiming at implementing practical applications of the QKD technologies \cite{peev2009secoqc,poppe2008outline}. Moreover, Tokyo QKD network was similar to the SECOQC in terms of infrastructure and was based on point-point links using trusted nodes topology. However, it included the implementation of the first Key Management Server (KMS) for centralised key management \cite{sasaki2011field}. The most recent network demonstration is the integrated space-to-ground quantum communication network in China \cite{chenintegrated}. This QKD network consisted of a long-distance fibre backbone network and two satellite–ground links and four quantum metropolitan area network where the backbone link covers over 2000km. A trusted node–free eight-user metropolitan quantum communication using a polarisation-entangled photon source has been implemented in the city of Bristol \cite{joshi2020trusted}.

Based on the several approaches to the coexistence of quantum and classical channels \cite{kumar2015coexistence, patel2012coexistence, eraerds2010quantum}, several field-trials were undertaken in different parts of the world. The Cambridge-Ipswich QKD network was one of these demonstrations in which a COW DV-QKD system in the O-band (IDQuantique Clavis3 system \cite{IDQCL3}) was integrated with 5x100~G classical DWDM channels in the C-band. Having the quantum and classical channels in different frequency bands significantly reduces any challenge in coexisting. The QKD network consisted of five nodes on a bus topology where the three intermediate nodes were used as trusted nodes \cite{wonfor2019field}. Another demonstration was the Cambridge Quantum Metro network consisting of coexistence between DV-QKD channels implementing a BB84 protocol with two decoy states and 2x100~G classical DWDM channels on a single link. The network consisted of 3 nodes acting as trusted nodes forming in a ring topology \cite{dynes2019cambridge}.

Classical and quantum channels coexistence highly benefit from intelligent control of resources, and an SDN controller is a perfect technology to accomplish that as shown in \cite{ou2018field}. SDN automation in QKD networks can be achieved based on real-time monitoring of quantum parameters such as SKR and QBER. Efforts to implement the first automated switching of the quantum channel in an SDN-controlled QKD network were reported in \cite{nejabati2019first,wang2019end}. In \cite{8657339}, a real-time monitoring of QKD parameters in an SDN-controlled QKD network was used to switch the quantum channel into a different optical fibre route under a physical-layer attack mitigation experiment. In the Madrid Quantum Network, the demonstration of a dynamic Software-defined CV-QKD quantum network was undertaken using commercial optical switches \cite{martin2019madrid,aguado2019engineering}. All of the previous mesh DV-QKD networks used trusted nodes or did not coexist quantum and classical channels in the same fibre. A trusted node is an intermediate node between to QKD links containing a QKD device from each link, that performs a relayed function between the two QKD links to establish end-to-end QKD connection. At the absence of quantum repeaters, these nodes are used to extend the reach of the QKD system and to enable the distribution of secure keys between the end points. Having a pair of QKD devices allow the quantum generated keys to be extracted and XORed with the keys of the QKD system in the next node. In that sense, trusted nodes consume a sender (Alice) and receiver (Bob) hardware at each intermediate nodes which add the cost prohibitively. Although trusted nodes add to the cost of the network, they are crucial to enable long distance end-to-end QKD operation as shown in \cite{wonfor2019field} where three trusted nodes were used for an end-to-end QKD operation using commercial QKD devices for over 120~km. Furthermore, since all the nodes in the network including intermediate trusted nodes are assumed to be safe from eavesdropping, any access to a trusted node cause a potential security risk and relinquish the strong security offered by quantum cryptography \cite{Zhang:18}. Therefore, a trusted-node free network that relies on optical switches does not only reduce the cost significantly relaxing the use of two QKD devices, but also does not stress the security requirements. To this end, we have recently demonstrated the concept of dynamic QKD as end-to-end quantum-secured inter-domain 5G implementation over a dynamically-switched optical network \cite{wang2019end} employing specially designed low-loss Q-ROADMs as the switching elements. On the application side a 5G network slicing with QKD using a SDN orchestrator has been also verified in \cite{wright20205g}. To further investigate the dynamic QKD approach in deployed optical fibre networks we have demonstrated a first field trial of a dynamic QKD networking capabilities across four nodes in the 5GUK testbed in Bristol \cite{tessinari2019field}. 

In this paper, we present a detailed study of our work on the first dynamic DV-QKD network in a field trial (in 5GUK deployed fibre in Bristol) and extend the analysis with a laboratory-based testbed to fully characterise the network performance in terms of quantum impairments considering the coexistence of carrier-grade classic channels and a quantum channel over switched network. Coexistence over multiple-hops using OXC switches and a comparison between different switching scenario are also presented. Additional findings of the coexistence over the deployed links of the 5GUK network are also evaluated. Moreover, new results of the SDN control implementation are also included in this communication. This paper is organised as follows: in Section \ref{sec:network_architecture} we discuss the overall network architecture. In Section \ref{sec:data-plane} we discuss the experimental system setup for the dynamic QKD network laboratory testbed followed by Section \ref{sec:control-plane} where we discuss the SDN control plane implementation. In Section \ref{sec:results} we present the results of the data and control plane and finally, in Section \ref{sec:conclusion} we conclude the paper.

\begin{figure*}[h]
    \centering
    \includegraphics[width=\linewidth,clip]{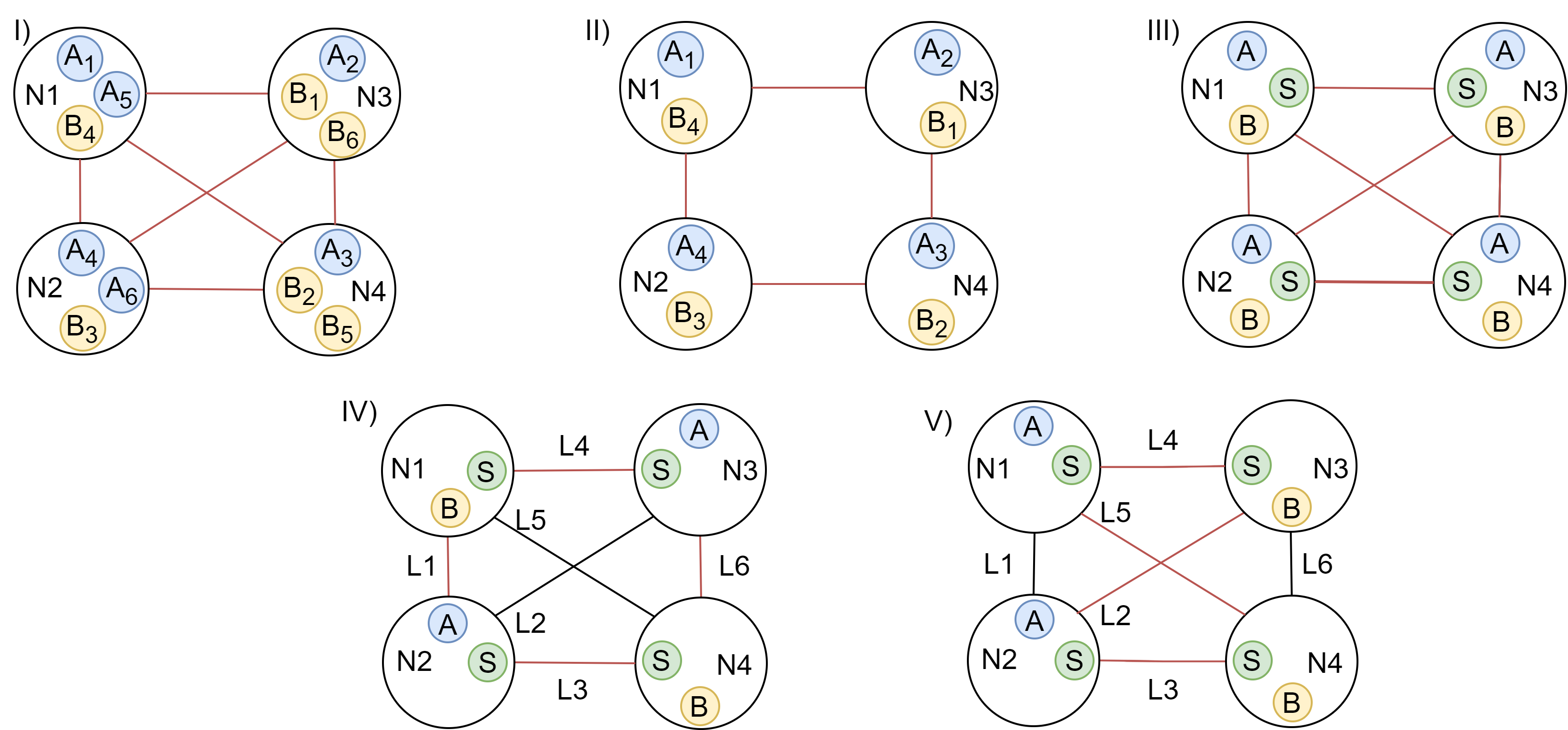}
    \caption{QKD Network Typologies. I) 6 QKD pairs trusted-node-free static configuration, II) 4 QKD pairs trusted-node static configuration, III) 4 QKD pairs trusted-node-free dynamic full-mesh configuration, IV) Trusted-node-free dynamic mesh configuration one, V) Trusted-node-free dynamic mesh configuration two. A: Alice, B: Bob, S: Optical Switch, Red line: Fibre with QKD communication, Black lines: Fibre without QKD communication.}
    \label{fig:top}
\end{figure*}

\section{Network Architecture}
\label{sec:network_architecture}

This section describes the overall network concept used in this paper. Full-mesh connectivity of a network with N nodes requires a minimum of $N(N-1)/2$ links. Therefore, as shown in \Cref{fig:top} I) to implement a direct QKD connection between any two nodes and avoid the relayed function of trusted node configuration, six pairs of QKD devices are required to cover all possible six links L1-L6. Of course, using only four pairs of QKD devices as shown in \Cref{fig:top} II) the six links could be covered by realising relayed nodes (trusted nodes) at the expense of using two QKD pairs to establish some links (e.g. node N2 to node N3 employs QKD pairs $A_4 - B_4$ and $A_1 - B_1$ and node N1 to node N4 using $A_1 - B_1$ and $A_2 - B_2$). \Cref{fig:top} III) shows a 4 nodes implementation of our dynamic trusted-node-free full-mesh QKD network where each node has an Alice (A), a Bob (B) and an optical switch (S). In this case, the Alice and Bob devices employed in the nodes are not dedicated to a specific Alice or Bob on any other node as the switch allows to physically connect the output of any Alice to the input of any Bob of the network. In this way, any node can establish a direct QKD link with any other node without using a trusted node due to the ability to switch the optical cross-connect port to the required destination port. The main advantage of the proposed switched QKD configuration is that the number of QKD pairs required scales linearly with the number of nodes (N QKD pairs for N nodes) for offering direct QKD links between any two nodes of the network, as opposed to $N(N-1)/2$ scaling without switches. In this experiment although the network has a full-mesh physical connectivity, only four QKD devices were used (two Alice devices and two Bob devices) and each node had either an Alice or a Bob depending on the configuration resulting in mesh QKD topology. \Cref{fig:top} IV) and \Cref{fig:top} V) show the two dynamic QKD configurations that were used in this experiment that when combined the cover all 6 links of the network. 

\begin{figure*}[t]
    \centering
    \includegraphics[width=\linewidth,clip]{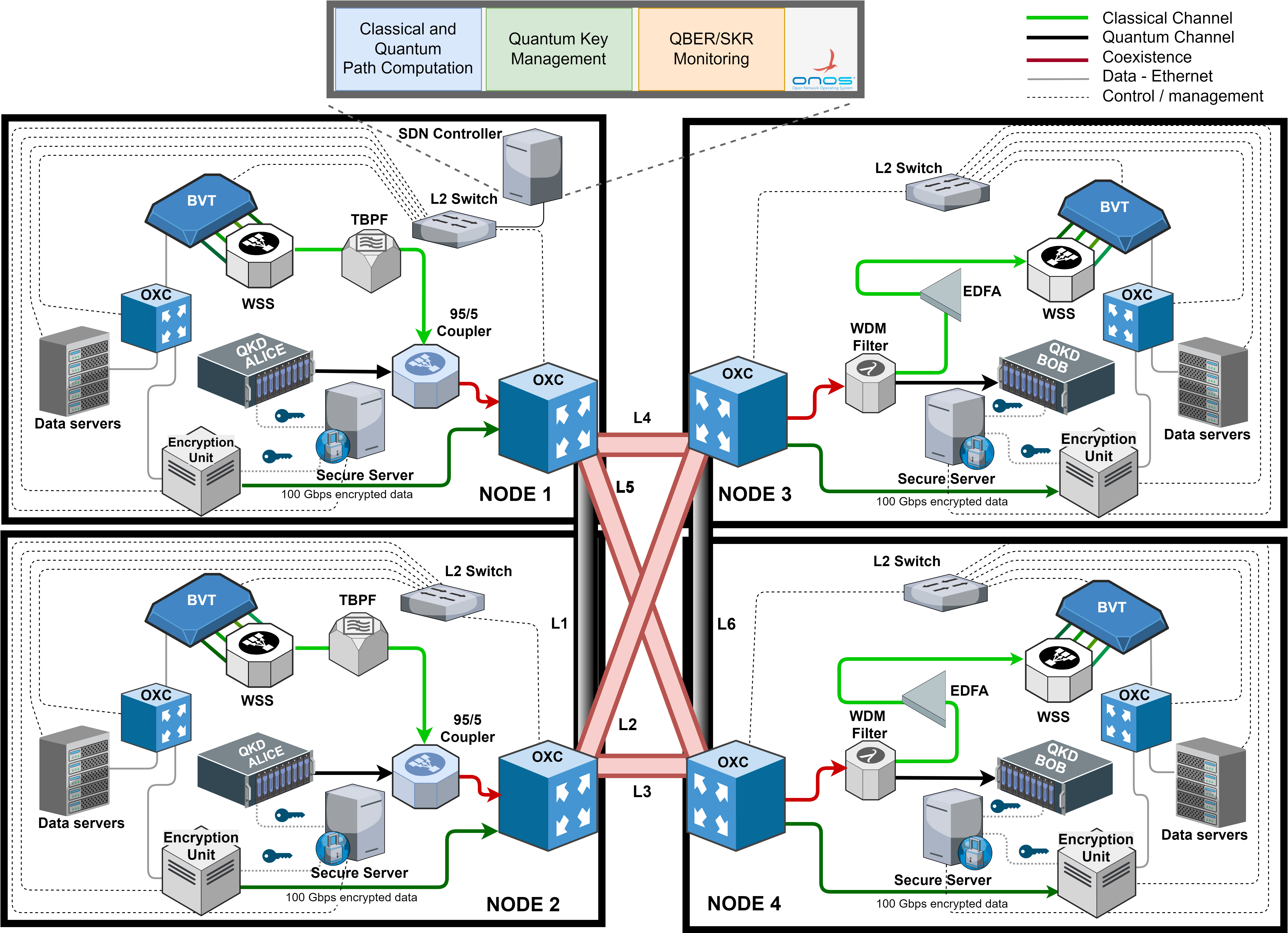}
    \caption{Trusted-node-free Dynamic QKD Network configuration two testbed. Red link: Fibre with QKD communication, Black link: Fibre without QKD communication.}
    \label{fig:network}
\end{figure*}

As shown in \Cref{fig:network}, the network is divided into control and data planes. The control plane contains the SDN controller and the required connectivity to the data plane equipment whereas the data plane represents the optical fibre infrastructure, optical equipment and QKD-related devices. The SDN controller is responsible for the computation, creation, and management of the complete path that traverses optical and SDN switches between the nodes. The SDN Controller is individually controlling the switches, QKD terminals, and encryption/application server in each node to create secure channels. The SDN controller utilises a quantum aware path computation mechanism, that calculates the best path for QKD and classical channel including power and losses, for minimal effect on the quantum channel. The SDN controller is also responsible for the establishment and management of paths that traverses between the nodes, including both classical and quantum channels and requires an L2 network to communicate with the data plane devices as demonstrated in \cite{tessinari2021demonstration}. 

To be able to perform the aforementioned tasks, the controller requires some essential components, such as databases to store network data (e.g. routing and connection tables), driver modules needed to control multiple different vendor devices, communication modules that understand different protocols (e.g. gRPC and Netconf), and finally intelligent algorithms that could use both heuristics or machine learning to optimise the usage of the network resources.   
The computational units host the software needed to communicate and control the equipment, including the Key Management System (KMS). The KMS is an entity that manages keys in a network in cooperation with one or more other KMS. When new cryptographic keys are generated by a QKD device, the keys are securely stored in a database (also called as key store) managed by the KMS. Such key stores provide a "key buffer" within the network, effectively decoupling the key generation process from the key consumption applications, allowing greater tolerance to bursts of key usage as also to temporary unavailability of the key-generation devices. Additionally, the KMS is responsible for monitoring and recording the usage and generation of keys, thus providing valuable statistics that can be used by the SDN controller. In the next section we describe the implementation of this architecture, as we describe the experimental setup.

Next, on the data plane, each node is composed of several components, including but not restricted to, optical switches, transponders, QKD equipment, data encryptors, computational servers, and filtering devices. Sharp filters are required to partly isolate the noise generate from the high power classical channels (orders of magnitude higher than quantum channels) to provide a sufficient isolation for the quantum  channel; hence improving its performance.

\section{Experimental System Setup}
\label{sec:testbed}

\subsection{Data Plane}
\label{sec:data-plane}
\Cref{fig:network} shows a details testbed setup of \Cref{fig:top} V. This testbed is used to demonstrate the dynamic QKD-networking using two QKD pairs. The network comprises four nodes interconnected via standard multiple single-mode fibre (SSMF) links resulting in a full-Mesh classical network due to physical fibre connecting all four nodes simultaneously. Each node is equipped with optical cross-connects (OXC), a bandwidth variable transponder (BVT), a QKD device, an encryption unit, data servers and a secure server. The testbed was designed to provide flexibility enabled by the SDN Controller. It is possible to transmit classical data using the BVT, coexisting or not with the quantum channel, while it is also possible to transmit encrypted data. This section describes the testbed in details.


The BVT includes four ports at 100~Gbps (25~Gbaud) data rate each with PM-QPSK modulation format. Each of the ports can be tuned to any of the 100 wavelengths in the C-band included in the ITU-T grid with 50~GHz offset. The QKD unit used is a commercially available IDQ Clavis2 DV-QKD \cite{RevModPhys.74.145} system with an auto compensating interferometric setup and quantum random number generators (QRNGs) to create secret keys. These Clavis 2 systems support fully automated sifting for the BB84 protocols and key distillation. 
The last equipment included in each node is a software encryption unit Dell PowerEdge server, responsible for the software encryption and used to secure the user data transmission with $AES-256$ encryption.      
\par As shown in \Cref{fig:network} in Node 1 and Node 2, the four coherent output ports of the BVT are coupled using a $1\times4$, 6-dB coupler, for a total throughput of 400~Gbps. A controllable q-ROADM is used to enable classical data channels and QKD signal routing and switching functionality. The q-ROADM provides low loss switching capability for the quantum channel that has a 10~dB limited power budget, as this consists of the filtering stages and the OXC devices. Moreover, the q-ROADM allows the dynamic reconfiguration of a hybrid QKD-classical network by allowing the arbitrary multiplexing of classical wavelengths and quantum channels at any port (or any degree) of the q-ROADM \cite{wang2019end}. The output is again coupled via a 95/5 ratio and insertion loss of 13~dB for the 5\% port used to enable low power loss in the quantum channel. The second port of this 95/5 coupler is used to exchange the encoded photons of the discrete-variable DV-QKD Alice unit considering a power loss of less than 0.5~dB for the 95\% port. The output of the 95/5 coupler is connected to the OXC and the coexisting QKD and classical signals are injected into the optical link via the suitable cross-connection. The OXC used is a SDN-enabled optical fibre switch with typical optical losses per cross-connection of 1~dB. 
In these nodes, the encryption server will interface to the IDQ Clavis 2 unit for the QKD protocol and an optical output of 100~Gbps data rate \cite{tessinari2021demonstration,arabul2022100} will transport the encrypted data towards the optical link via the OXC. In Node 3 and Node 4, the OXC will cross-connect the incoming coexisting signal to an optical band pass and rejection (stop) filter (BPRF) with 0.8~dB of loss for the band pass port, before connecting to the the IDQ Clavis 2 Bob unit. The band pass port of the BPRF has an optical bandwidth of 100~GHz centered at the 1551.7~nm wavelength of the QKD units. For the rejection port of the BPRF, the quantum channel is blocked and the combined classical signals are optically amplified by an erbium-doped fibre amplifier (EDFA) to boost the optical power to an acceptable level for detection. The output of the EDFA feeds a $1\times4$, 6~dB splitter and its four different optical ports are each connected to a coherent receiver of the BVT. Finally, in these nodes, the decryption server will extract the data received after processing the key. Table \ref{tab:parameters} summarises the main parameters of the implemented fully-meshed QKD dynamic network.



\begin{table}[t]
\centering
\caption{Parameters for Dynamic QKD Networking Testbed}
\label{tab:parameters}
\begin{tabular}{cc}
\hline
\hline
\multicolumn{1}{c}{Parameters}  &   \multicolumn{1}{c}{Value}       \\
\hline
\multicolumn{2}{c}{\textit{Classical Channels}} \\
Number of Channels  &   4   \\

\begin{tabular}[c]{@{}l@{}}Classical Channel\\ \ \ \ Wavelengths\end{tabular}         & \begin{tabular}[c]{@{}l@{}}1550.52 nm, 1550.12 nm,\\ 1549.72 nm, 1549.32 nm\end{tabular} \\

\begin{tabular}[c]{@{}l@{}}Classical Channel\\ \ \ \   Frequencies\end{tabular}         & \begin{tabular}[c]{@{}l@{}}193.35 THz, 193.40 THz,\\ 193.45 THz, 193.50 THz\end{tabular} \\

Grid Spacing        &   50 GHz      \\
Modulation Format   &   PM-QPSK     \\
\begin{tabular}[c]{@{}l@{}}Optical Signal-to-Noise\\  \ \ \ \ \ Ration (OSNR)\end{tabular} & 20 dB \\
Capacity per Channel& 100 Gbps      \\
Total Capacity      &   400 Gbps    \\
Pre-FEC Level       &   15\%        \\
Detector sensitivity*& -35 dBm      \\
                                    \\
\multicolumn{2}{c}{\textit{Quantum Channel}}    \\
DV-QKD Wavelength   & 1551.7 nm (C-band)    \\
DV-QKD Frequency   & 193.20 THz    \\
QKD Protocol        & BB84          \\
Maximum Distance    & 50 km @10 dB loss \\
                                    \\
\multicolumn{2}{c}{\textit{EDFA}}   \\
Noise Figure    & 5 dB              \\
Operation Mode  & Continuous optical power  \\
                                    \\
\multicolumn{2}{c}{\textit{Encryption/Decryption}}  \\
Encryption technique    & AES-256   \\
Encryption data rate    & 10 Gbps   \\
                                    \\
\multicolumn{2}{c}{\textit{Optical band pass/Rejection Filter (OBRF)}}   \\
\begin{tabular}[c]{@{}l@{}}Insertion loss \\ band pass port\end{tabular}     & 0.5 dB    \\
\begin{tabular}[c]{@{}l@{}}Center wavelength\\ \ \ \ band pass port\end{tabular}   & 1551.7 nm \\
Bandwidth band pass port & 100 GHz   \\
\hline
\hline
\multicolumn{2}{l}{\textit{*Corresponding to PM-QPSK Modulation @100 Gbps and back-to-back}}  \\
\end{tabular}
\end{table}

\subsection{Control Plane}
\label{sec:control-plane}
As \Cref{fig:network} shows, each node has several components that need to be monitored and controlled by the SDN controller. To provide connectivity between the SDN controller and all the controllable devices in our deployed network we devised a supporting management network. One particular component is the Secure Server, that connects to L2 switches and other Secure Servers (using 10 Gbps small form-factor pluggables (SPFs)), thus accounting for the control plane logical and electrical connections described in \Cref{fig:network}.

The Secure Servers also host the Key Manager System (KMS) that doubles down as a key database and as an agent to manage and control the QKD devices. The KMS software of choice was the CQP Toolkit, developed in-house \cite{cqpToolkit}. Additionally, the Secure Servers also hosts the OXC agent and the encryption unit agent.

As can be seen in \Cref{fig:network}, Node 1 also hosts the SDN controller used in our testbed, in addition to all the described components. The SDN controller is responsible for the establishment and management of the complete path that traverses the optical SDN switches between the nodes. It can handle paths for the quantum channel, classical channels, and coexisting QKD and Classical channels. Additionally to the switches, the QKD terminals and the encryption units are also controlled.

In the current controller state, the routes are calculated offline using the network information and then uploaded to the controller. This network information includes the OXC ports, link connections and losses, and available routes. Therefore, the path assignment consists of selecting the most suited route available for the incoming request (e.g. fibres with smaller losses for a quantum channel). Classical channels can be established between any two nodes and quantum channels between any two matching devices (i.e. one Alice and one Bob) as per the user’s request.

The controller monitors the SKR and QBER in real-time and fetches these parameters from the QKD agent (CQP Toolkit) using gRPC whenever the parameters are updated (every two minutes). The controller also monitors classical channels parameters such as BER and the number of errors. These parameters are updated every second, however, the controller fetches the classical parameters when a new set of quantum parameters (QBER and SKR) is generated by the QKD devices. Furthermore, when a new QKD connection is established the configurations table is automatically updated based on the available links. Although the switches configuration occurs instantly, establishing a new QKD link requires 10-15mins to authenticate the QKD devices and generate keys.

\begin{figure}[t]
    \centering
    \includegraphics[width=\linewidth,clip]{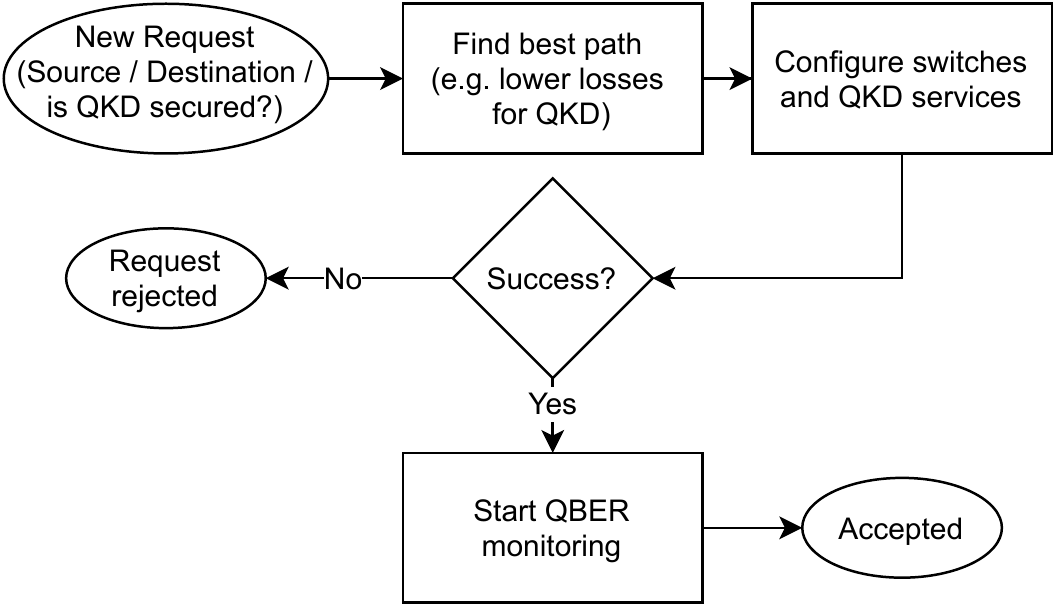}
    \caption{New request flowchart. In addition to classical resource allocation, the SDN controller also considers QKD-impacting characteristics, such as link losses, and optimise the allocation to provide a better QKD experience when a QKD-secured connection is requested.}
    \label{fig:sdnflow}
\end{figure}

\Cref{fig:sdnflow} summarises the flow when a new request arrives to the SDN controller.
The start connection request identifies a source site and a destination site and retrieves all the information pertaining to that particular pair from the routing table. This information includes the cross connections (e.g. OXC's input and output ports) needed to fulfil the path between sites. The SDN controller utilises the information to install the flow rules in the OXC. After the OXC devices are configured, the controller issues the start QKD message towards to the KMS agents, triggering the start of the key generation process. Thereafter, the monitoring of the Quantum Bit Error Rate (QBER) starts, raising an event within the controller in case there is a deviation the QBER values. If the QBER is below a QBER  threshold (6\% by default), it is deemed as acceptable and the encryption unit is set to initiate the encryption software, creating an encrypted tunnel between the endpoints. However, if at any moment the QBER monitored is above the threshold, the current quantum channel quality is noted as unsatisfactory and therefore the controller proceeds to fetch the next available route in the routing table connection and the aforementioned process starts again.

Once the whole system is running, the quantum keys are generated by the IDQClavis2 and stored on databases managed by the KMS system. In addition to the KMS, CQPToolkit software suite also encompasses an interface to control the Clavis devices and an encryption application. This controlling interface allows the SDN controller to start/stop IDQClavis2 devices and subsequently to request statistics to monitor the quantum channel parameters. All communication between CQPToolkit and the SDN controller is done using gRPC as communication protocol. As encryption application, CQPToolkit suite ships with the QTunnelServer software which implements one solution to the problem of sending encrypted data from one site to another. It creates an encrypted tunnel thus allowing secured data transfer by using AES-256 encryption.

\section{Results}
\label{sec:results}

\subsection{Results of the Fully-Meshed Dynamic Network without Coexistence}
\Cref{tab:QKD_parameters} shows the main parameters of both configuration of the dynamic QKD network of \Cref{fig:top} IV and V. The parameters of the optical fibre links are described such as fibre lengths and end-to-end power losses. Also, \Cref{tab:QKD_parameters} depicts the number of cross-connections per link required for dynamically switch to the quantum channel requested. The quantum parameters of quantum bit error rate (QBER) and secret key rate (SKR) are represented in this table without coexisting with classical channels. In addition, the combination of links for multihop scenarios is included to show the performance of the dynamic QKD network without coexistence. As observed, the link L1 with the lowest power budget (5.19~dB) excluding back-to-back configuration achieves the lowest QBER of 1.31\% and highest SKR of 1762.06~bps. However, since L6 has the highest power budget of 9.61~dB and two cross connections, results show the highest QBER of 3.65\% and the lowest SKR of 360.08~bps, being near the power budget limit of the QKD system used. L5 and L6 are field deployed fibres that were purposely chosen to reach the power budget limit of the QKD system. In \Cref{tab:QKD_parameters} $A_2 - B_1$ means the direct connection from Alice in Node 2 to Bob in node 1 whereas L1 + L2 is the connection from Bob in node 1 to Alice in node 3 via the switch in node 2 as shown in \Cref{fig:top} IV.

\begin{table}[h]
\centering
\caption{Dynamic QKD network parameters}
\label{tab:QKD_parameters}
\begin{tabular}{@{}cccccccc@{}}
\hline
\hline
Link    &
\begin{tabular}[c]{@{}c@{}c@{}}fibre\\Length\\ (km)\end{tabular} &
\begin{tabular}[c]{@{}c@{}c@{}}End-to-\\ End Power\\ Budget(dB)\end{tabular} &
\# OXC &
\begin{tabular}[c]{@{}c@{}}QBER\\(\%)\end{tabular} &
\begin{tabular}[c]{@{}c@{}}SKR\\(bps)\end{tabular} &
\\ \hline
 Back-to-Back & 0 & 4.99 & 1 & 1.02 & 2575.69  \\
 L1 $(A_2 - B_1)$& 0.5 & 5.19 & 2 & 1.31 & 1762.06  \\
 L2 $(A_2 - B_3)$& 1 & 5.70 & 2 & 1.43 & 1414.15  \\
 L3 $(A_2 - B_4)$& 5.8 & 6.90 & 2 & 1.70 & 1078.03  \\ 
 L4 $(A_3 - B_1)$& 4.7 & 7.00 & 2 & 1.91 & 895.54  \\
 L5 $(A_1 - B_4)$& 1.625 & 9.22 & 2 & 3.03 & 418.74  \\
 L6 $(A_3 - B_4)$& 1.625 & 9.61 & 2 & 3.65 & 360.08  \\
 L1 + L2 & 1.5 & 7.44 & 3 & 2.01 & 819.22  \\
 L1 + L3 & 6.3 & 8.60 & 3 & 2.70 &  533.97 \\
 L1 + L4 & 5.2 & 8.79 & 3 & 2.73 & 542.26 \\ 
 L2 + L3 & 6.8 & 9.14 & 3 & 3.03 &  430.56 \\
 L2 + L4 & 5.7 & 9.42 & 3 &  3.05 &  417.94 \\
\hline

\end{tabular}
\end{table}

{\newcommand\figureSize{0.475}
\begin{figure*}[!h]
    \centering
    \begin{tabular}{cc}
        \subfloat { \includegraphics[width=\figureSize\linewidth]{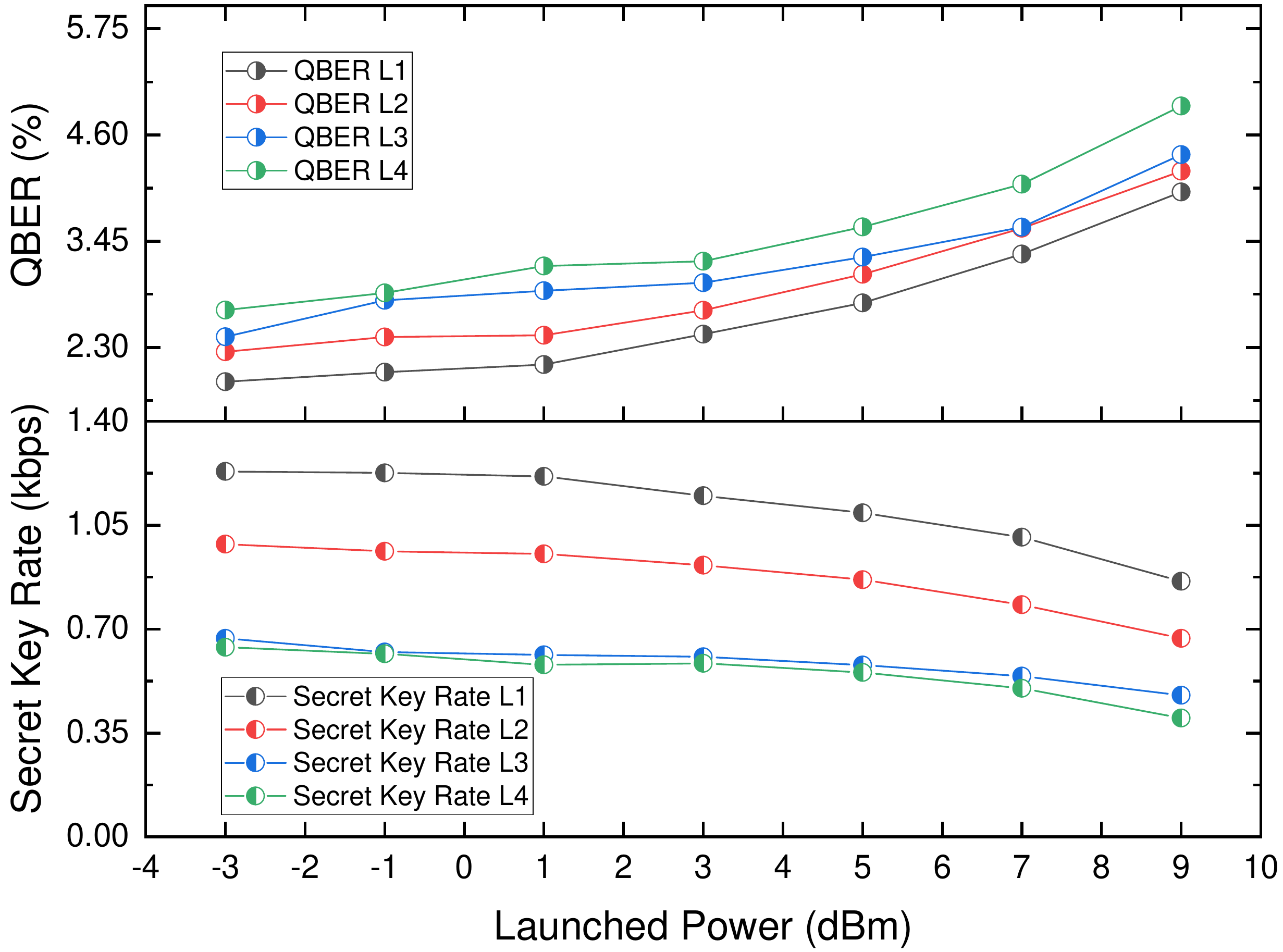}             \label{fig:1ClCh} } & 
        \subfloat { \includegraphics[width=\figureSize\linewidth]{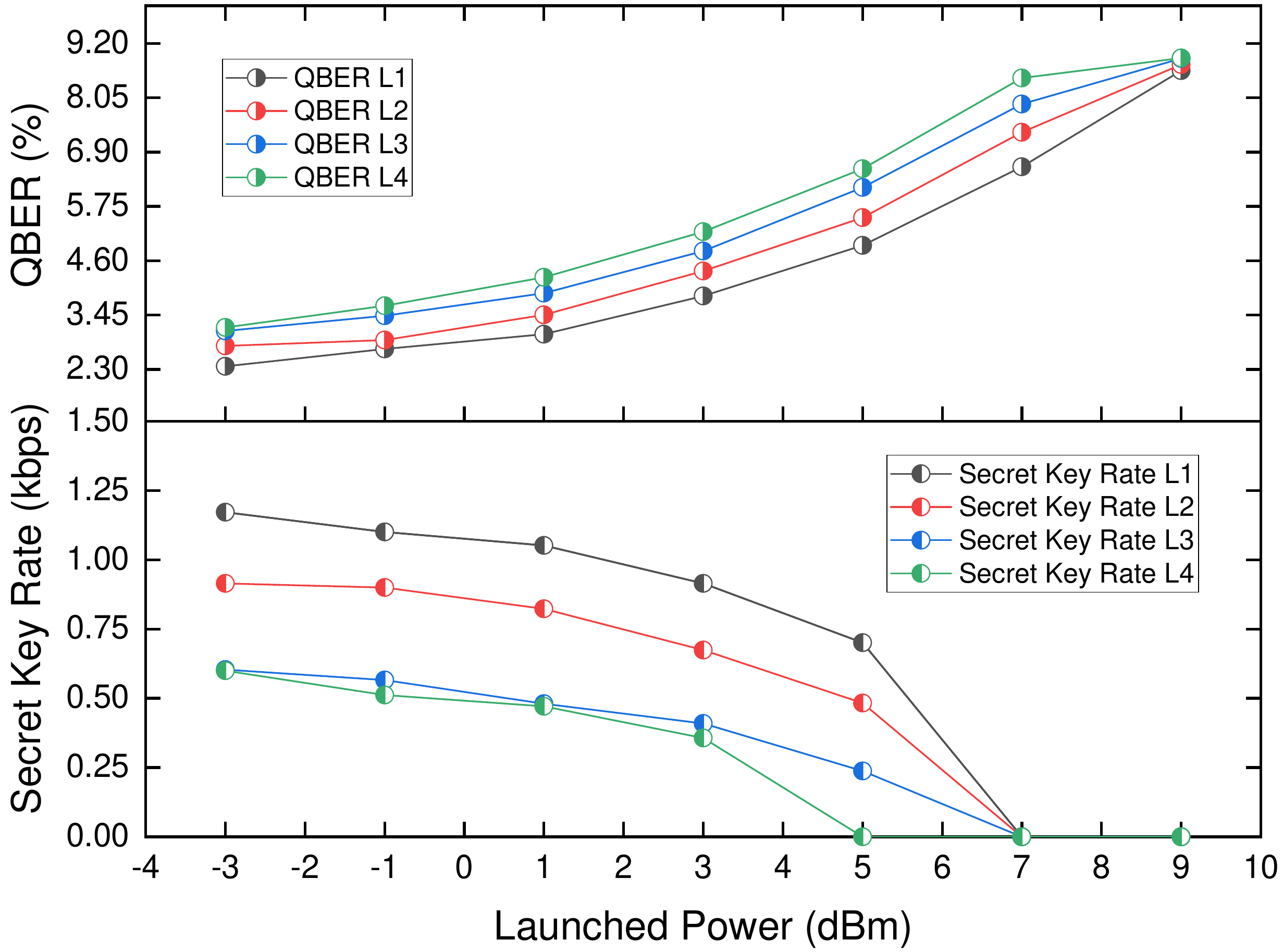} 
        \label{fig:4ClCh}
        } \\
        (a) & (b) \\ 
        \subfloat { \includegraphics[width=\figureSize\linewidth]{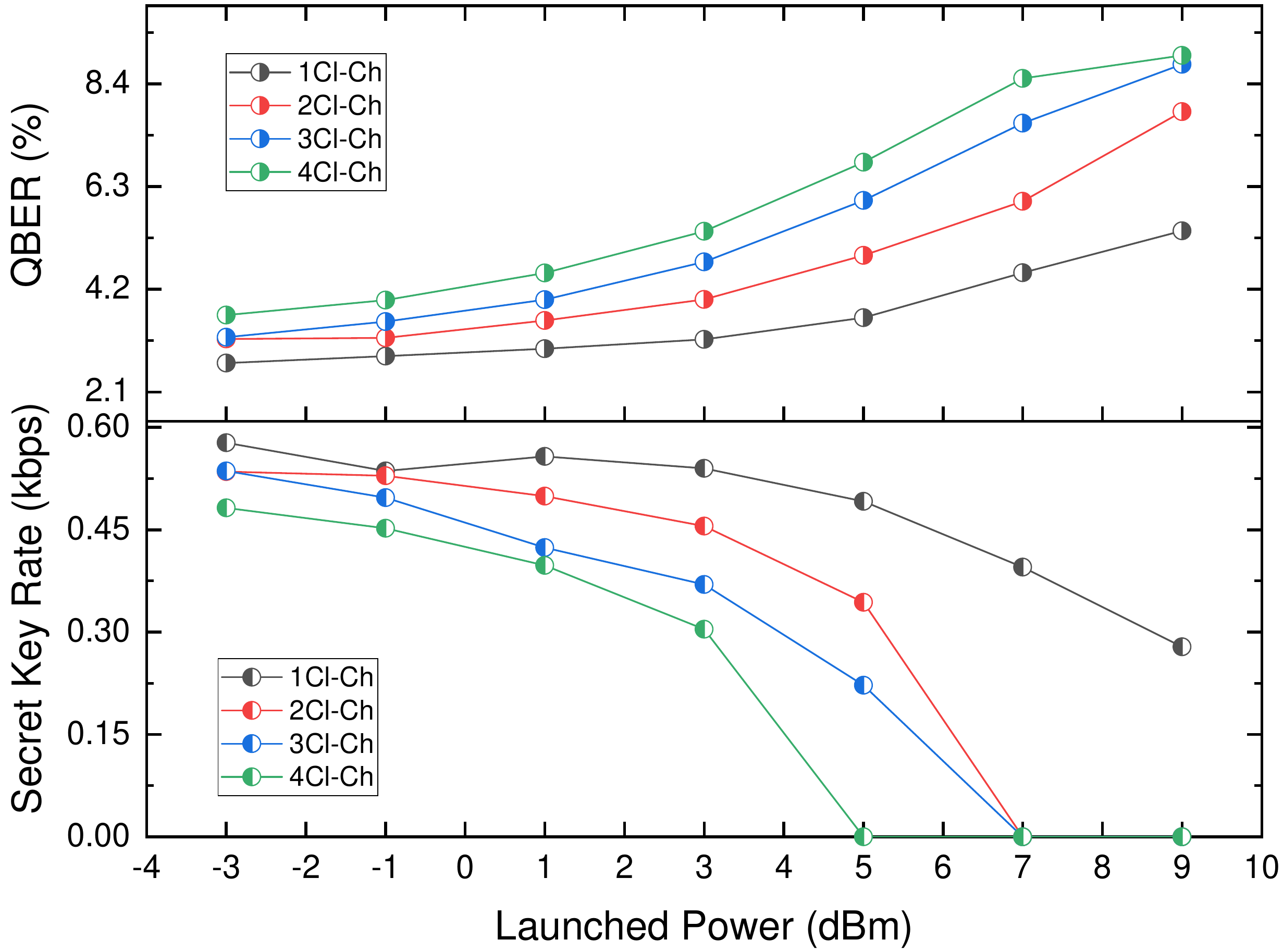}             \label{fig:L1+L2} } &
        \subfloat { \includegraphics[width=\figureSize\linewidth]{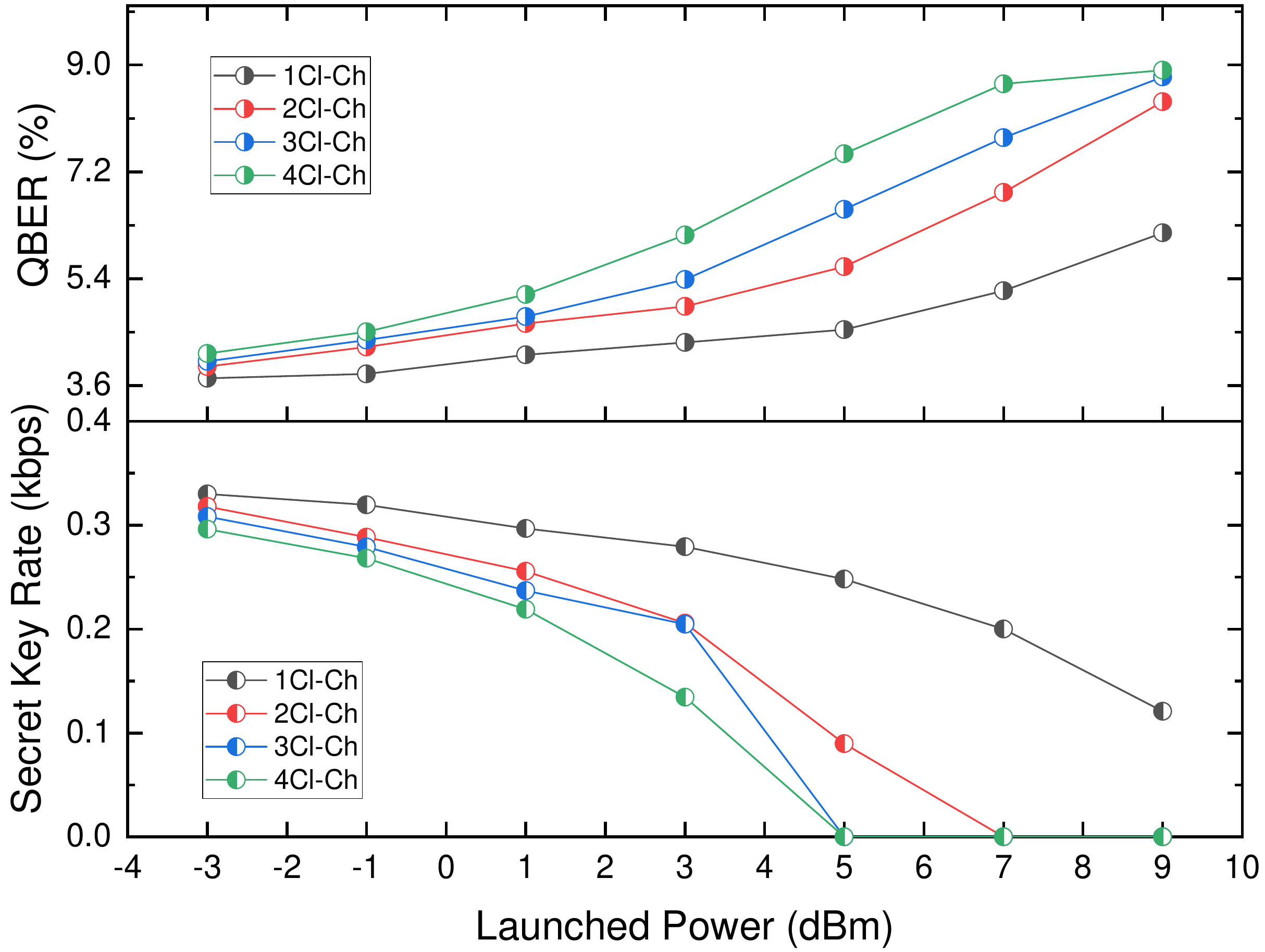}       \label{fig:L1+L3} } \\
         (c) & (d) \\ 
    \end{tabular}
    \caption{
    (a) Coexistence of quantum channel and one classical channel ($F_c 193.35\ THz$) for four different links. 
    (b) Coexistence of quantum channel and four classical channels for four different links.
    (c) Coexistence of quantum channel and four classical channels for link L1+L2 
    (d) Coexistence of quantum channel and four classical channels for link L1+L3
    }
    \label{fig:All_Charts}
\end{figure*}
}
{\newcommand\figureSize{0.475}
\begin{figure*}[!t]
    \centering
    \begin{tabular}{cc}
        \subfloat { \includegraphics[width=\figureSize\linewidth]{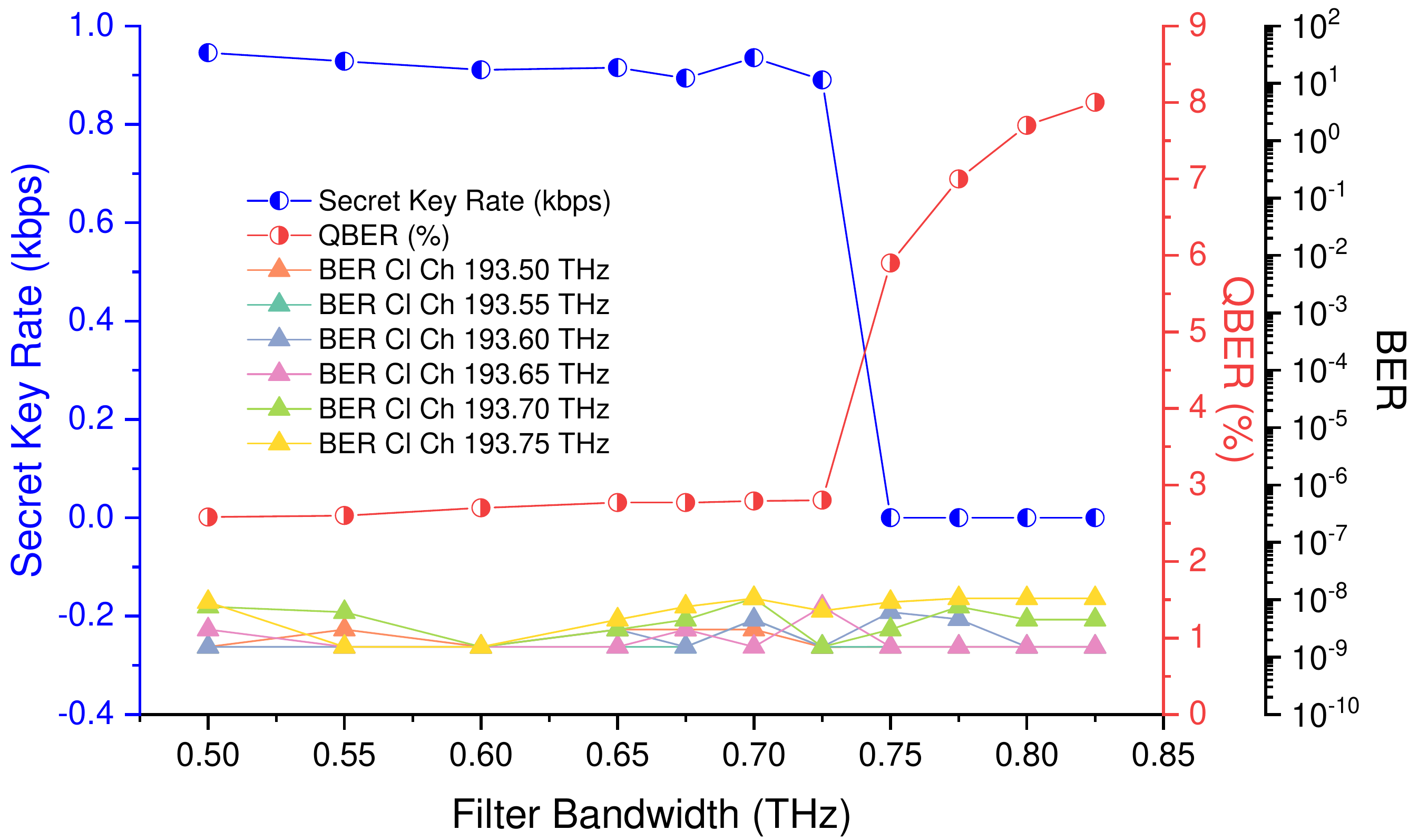}             \label{fig:HPN-WTC} } & 
        \subfloat { \includegraphics[width=\figureSize\linewidth]{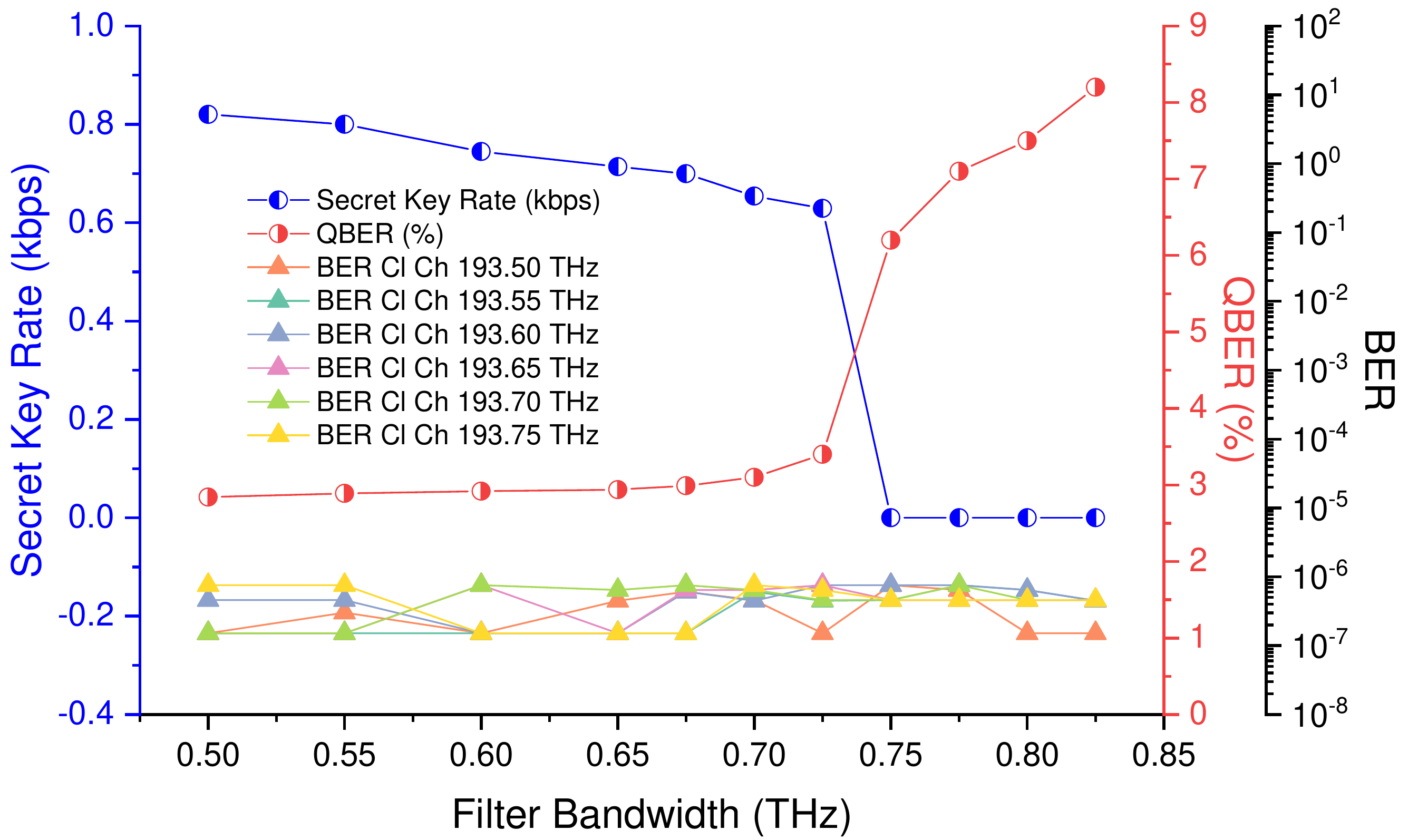} 
        \label{fig:NSQI-WTC}
        } \\
        (a) & (b) \\ 
    \end{tabular}
    \caption{
    (a) Coexistence of quantum channel and six classical channels in a 1.9~km field-deployed fibre (HPC-WTC)
    (b) Coexistence of quantum channel and six classical channels in a 2.7~km field-deployed fibre (NSQI-WTC)
    }
    \label{fig:testbed_NSQIHPN_WTC}
\end{figure*}
}

\subsection{Results of the Fully-Meshed Dynamic Network with Coexistence}
To investigate the effect of Raman noise over the DV-QKD channel, the testbed of \Cref{fig:network} is used considering links L1, L2, L3 and L4 and one classical channel centered at the frequency of 193.35~THz with 150~GHz spacing from the quantum channel. \Cref{fig:1ClCh} shows the measured SKR and QBER of the quantum channel at different launched optical power levels of the classical channel. As observed, when the launching power is more than 1~dBm, the QBER and SKR deteriorate due to the noise leakage into the 100~GHz bandwidth of the internal filter of the Bob DV-QKD unit. Moreover, the close QBER values achieved by the DV-QKD channel over all the links reflect the small variation of optical attenuation of 2~dB for each link. In addition, at the BVT highest possible launch power of 9~dB, the lowest SKR achieved is 350~bps with continuous key generation for link L4. As shown in \ref{tab:QKD_parameters}, L3 is longer than L4 and therefore, the Raman noise should deteriorate the QKD performance for L3 more than L4 due to the longer fibre. However, as shown in \Cref{fig:1ClCh} the SKR is identical for both L3 and L4. This is due to the short fibre length (<6km) of both links, the low coexistence power and the almost equivalent end-to-end power budget for both links.

\Cref{fig:4ClCh} shows the impact of coexisting four classical channels (193.35~THz, 193.40~THz, 193.45~THz and 193.50~THz) over the quantum channel (193.20~THz). As observed from \Cref{fig:4ClCh}, the quantum channel deteriorates faster compared to \Cref{fig:1ClCh} due to the combination of Raman noise and other nonlinearities. Considering the two frequencies 193.35~THz and 193.50~THz as $f_1$ and $f_2$ respectively, one product of the four wave mixing would be  $f_3 = 2f_1 - f_2 = 193.20\ THz$ which is the quantum channel frequency which degrade the performance of the quantum channel due to the additional noise from such phenomena. Therefore, more noise leakage into the Bob DV-QKD unit occurs due to higher aggregated launch power (16~dB when transmitting four classical channels at 9~dB), higher Raman noise and other nonlinear effects such as four wave mixing. It also observed that at a launch power of 7~dB per channel, the QBER values exceed the threshold of 6\% causing the SKR to be zero bps. 

\Cref{fig:L1+L2} and \Cref{fig:L1+L3} show the SKR and QBER over the combined links of L1+L2 and L1+L3. In both figures, increasing the number of classical channels is presented to reflect the impact of incremental channels over the quantum channel. As observed, similar trends of QBER and SKR deterioration appear for both cases of links due to the optical power budgets in the vicinity of 8~dB. Also, it is clear in both links that by adding a classical channel, the total aggregated power is increased and the Raman noise effect is proportional to the power added, worsening the quantum channel performance. It can also be observed that the QBER values exceed the threshold of 6\% causing the SKR to be zero bps at a launch power of 5~dB per channel which is different comparing to \Cref{fig:4ClCh}. This is due the additional losses of links L1+L2 and L1+L3 and the additional cross-connection in the optical switch as shown in \Cref{tab:QKD_parameters}.

\subsection{Coexistence over Bristol City 5GUK Testbed}

To further explore the coexistence of a DV-QKD channel with classical carrier-grade channels, two links of the Bristol City 5GUK testbed were used \cite{tessinari2019field}. One link connects the site HPN to the WTC node, with a fibre length of 1.9~km total optical fibre attenuation of 4.68~dB. The other link interconnects the sites NSQI to the node WTC passing through the node HPN (2 hops) with a fibre length of 2.7~km total optical fibre attenuation of 5~dB. \Cref{fig:HPN-WTC} shows the QBER and SKR for the link HPN-WTC. A quantum DV-QKD channel coexisted with 6 PM-QPSK 100~Gbps channels with 50~GHz spectrum space difference between them (from 193.5~THz to 193.75~THz). To evaluate the performance of the coexistence over these channels, the bandwidth of the optical band pass filter of \Cref{sec:testbed} is tunned to gradually allow noise proliferation into the quantum channel. As observed, when the filter bandwidth is in the range of 500~GHz to 725~GHz, the SKR obtained is higher than 890~bps and the QBER is lower than 2.8\%. However, for filter bandwidths higher than 750~GHz, the noise leakage over the quantum channel will impede the key generation due to high QBERs of more than 5.9\% causing the SKR to plummet to zero bps. For the classical channels, the BER measured was $3.5\times10^{-9}$ average for the channels selected.
\Cref{fig:NSQI-WTC} shows the QBER and SKR curve with respect to the filter bandwidth for the link NSQI-WTC. Compared to the link HPN-WTC (\Cref{fig:HPN-WTC}), the SKR and QBER obtained degrade faster. This is due to the additional crossconnection for the extra hop in the link which increases the power budget of the link.

\section{Conclusion}
\label{sec:conclusion}
We have demonstrated for the first time a trused node free dynamic QKD networking implementation over a testbed that spanned across four optical nodes interconnected in mesh topology with short links between nodes emulating the case of a dense metropolitan region. The coexistence of a DV-QKD channel and $4\times100$~Gbps classical channel was successfully demonstrated over multiple links with the ability to switch between the links using a centralised SDN controller. For the coexistence over the longest link (L3) with 5.8km, a minimum QBER of 1.7\% and a maximum SKR of 1078~bps was demonstrated for the DV-QKD simultaneously with a minimum average pre-FEC BER of $1.28\times10^{-8}$ for the error free classical channels. Investigations also prove that when coexisting four classical channels with 150~GHz spacing from the quantum channel, a minimum launch power of 7~dB per channel is required to deteriorate the key generation process with a SKR of zero bps. Moreover, this work also demonstrated the coexistence of a quantum channel and six classical channels through a field-deployed fibre in the 5GUK Test Network. The key generation process is maintained when the bandwidth of the optical band pass filter which is centred at 193.625~THz and is lower than 750~GHz. When the filter bandwidths is tunned to higher than 750~GHz, the noise leakage over the quantum channel will impede the key generation due to high QBER of more than 6\% causing the SKR to plummet to zero bps.

\section*{Acknowledgement}
This work acknowledges the EU funded project UNIQORN (820474), the EPSRC $EP/T001011/1$: UK QHub for Quantum Communications, and Facebook for the Voyager switches utilised in the experiment. For the purpose of open access, the author(s) has applied a Creative Commons Attribution (CC BY) license to any Accepted Manuscript version arising.

\bibliographystyle{IEEEtran}
\bibliography{IEEEabrv,refs}









\end{document}